\documentclass{jltp}
\input epsf 
\newcommand{\E}{{\rm e}} 
\newcommand{\D}{{\rm d}} 
\newcommand{\I}{{\rm i}}

\title{Persistent Currents versus Phase Breaking in Mesoscopic Metallic Samples} 
\author{U. Eckern and P. Schwab}
\address{Institut f\"ur Physik, Universit\"at Augsburg,
86135 Augsburg, Germany
} 
\date{\today} 
\runninghead{U. Eckern and P. Schwab}{Persistent Currents versus Phase Breaking}
\begin{document}

\maketitle

\begin{abstract}
Persistent currents in mesoscopic normal metal rings represent, even a
decade after their first experimental observation, a challenge to both,
theorists and experimentalists. After giving a brief review of the
existing -- experimental and theoretical -- results, 
we concentrate on the (proposed) relationship of
the size of the persistent current to the phase breaking rate. In
particular, we consider effects induced by noise, scattering at
two-level systems, and magnetic impurities. 

PACS: 05.30Fk, 73.23.Ra
\end{abstract} 
The question of the magnetic properties of ring molecules and metallic
rings has a long history. In analogy to the Aharonov-Bohm effect,
i.e. the tuning of the interference pattern in a double slit experiment
through the enclosed flux, it can be expected that the magnetization, or
the corresponding ``persistent current'', has a flux periodic contribution.
In this sense, and for sufficiently low temperatures, metallic rings are
very similar to ring molecules. But how large is the effect for a mesoscopic,
disordered ring? Is it experimentally observable with today's  technology?
 
The first successful experimental investigations which showed that persistent currents exist
in mesoscopic rings have been performed at the beginning of the 1990s and were most important for
the advancement of the 
field\cite{buttiker85,levy90,chandrasekhar91,cheung89,schmid91,oppen91,altshuler91,ambegaokar90}.
We will first present selected aspects of the theoretical concepts developed at the time;
see, for example, Ref.\ \onlinecite{eckern95}
for more details. 
We then compare the theoretical results with the available experimental data. 
Since no clear picture emerges there is still room for improvement of the theory. In particular,
we report on recent attempts to relate the size of the persistent current to the phase 
braking rate.

\section{GENERAL ASPECTS OF THE THEORY}

In the following we consider, for the sake of simplicity, an idealized
situation where the width of the metal ring is so small compared to the
circumference $L$ that magnetic field penetration into the metal can be
neglected. Then the energy and the
thermodynamic potential, $K(\phi)$, of a given sample depend on the magnetic
flux, $\phi$, and not explicitly on the magnetic field.
As the persistent current $I(\phi)$ is an equilibrium
property,
it can be calculated by taking the derivative according to
\begin{equation}
I(\phi) = - {\partial K(\phi)\over \partial \phi}
.\end{equation}
Depending on the variables to be kept constant in the experiment,
$K(\phi)$ is given by $F(N,\phi)$ or $\Omega(\mu,\phi)$
for the canonical and the grand canonical ensemble, respectively. 
The periodicity of $K(\phi)$ with period $\phi_0 = h/e$ allows the Fourier
decomposition of the current,
\begin{equation}
I(\phi) = I_{h/e} \sin( 2\pi \phi/\phi_0) + I_{h/2e} \sin( 4 \pi \phi/\phi_0) + \cdots
.\end{equation}
For simplicity we do  not further discuss the subtle
questions concerning differences 
\cite{schmid91,oppen91,altshuler91} 
between 
$F(N,\phi)$ and 
$\Omega(\mu, \phi)$
and we concentrate on the grand
canonical ensemble.
Furthermore we will focus on diffusive rings, i.e. we include disorder with an elastic
mean free path $l$ such that $\lambda_F \ll l \ll L$.
We also assume that the circumference is much smaller than the localization length.

\section{NON-INTERACTING ELECTRONS}
For non-interacting electrons the complete thermodynamics is determined from the
single particle density of states.
In order to compute the persistent current, consider first the grand canonical 
potential
\begin{equation}
\Omega(\mu, \phi)= - 2 {\cal V} k_B T \int \D E N(E,\mu) \ln \{  1+ \exp[ -(E-\mu)/ k_BT ]\} 
\; ,\end{equation}
where the factor two is due to the spin, ${\cal V}$ is the volume, and $N(E)$ is the density
of states. The energy levels and hence the density of states depend on the magnetic flux,
which is to be calculated.

\begin{figure}
\hspace{1.0cm}{\epsfxsize=11.0cm\epsfbox{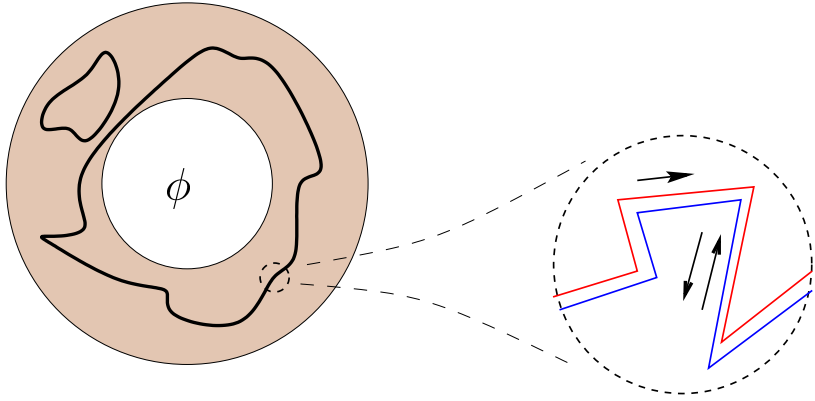}} 
\caption{Paths with winding number zero and one. The scaled-up part represents the
microscopic processes with amplitudes $A_i$ and $(A_j)^*$. $A_i$ and $A_j$
correspond to identical or time reversed paths.}
\label{fig1}
\end{figure}

In an ensemble of weakly disordered rings the disorder configuration will change
from ring to ring.
Accordingly the density of states and the persistent current will be statistically
distributed.
It is well known that the average density of states, $N_0 = \langle N(E) \rangle $,
is a flux independent quantity except for corrections which are exponentially small and
proportional to $\exp(-L/2l)$. Thus the average persistent current, 
as computed in the grand canonical ensemble, is negligibly small.
Fluctuations of the density of states, on the other hand, are not exponentially suppressed
and have to be considered. 
Manipulating the expression given in Ref.\ \onlinecite{altshuler87}
one arrives at
\begin{eqnarray}
\lefteqn{ \langle \delta N(E, \phi ) \delta N(E+ \hbar \omega, \phi' )\rangle
=  {L \over \pi^2 \hbar^2 {\cal V}^2 } 
\int_0^\infty \D t  \cos( \omega t ) t }\cr
\label{eq4}
&& \times \left[ P(0,t)+ \sum_{m, \pm} P(mL, t ) \cos( 2\pi m (\phi\pm \phi')/\phi_0 ) \right]
,\end{eqnarray}
where 
\begin{equation}
P(x,t)= {1\over \sqrt{ 4 \pi  D t}  } \exp \left(- {x^2 \over 4 D t} \right)
\end{equation}
is the solution of the diffusion equation.  
This means that the density of states fluctuations are proportional to the
integrated probability that a diffusing particle returns after time $t$.
For the ring structure there are topologically different possibilities 
to return to the origin, see Fig.\ \ref{fig1}.
Paths with zero winding number lead to the contribution $P(0,t)$ and do not depend on magnetic
flux, while paths which traverse the ring $m$ times lead to the term 
$P(mL, t)$ and are flux dependent. The reason is the following: First recall that a 
probability is in quantum mechanics always the product of
amplitudes, $P \sim (\sum A_i)( \sum A_j)^*$.
From all possible combinations $A_i^{\vphantom *} A_j^*$, Eq.\ (\ref{eq4}) then selects
pairs of equal paths $j=i$ and pairs of time reversed paths $j= {\bar i}$ as
illustrated in the figure.
In the presence of a magnetic flux the phase of
a closed path is shifted according to $A \to A \exp(2 \pi \I \phi/\phi_0)$, 
where $\phi $ is the enclosed magnetic flux.
For a pair of two equal paths, but in the presence of two different
magnetic fluxes $\phi$ and $\phi'$, one obtains
$ A_i^{\vphantom *} A_i^* \to A_i^{\vphantom *} A_i^* \exp[ 2 \pi \I (\phi - \phi')]$. 
For a pair of time reversed paths,
on the other hand, the result is
$A_i^{\vphantom *} A_{\bar i}^* \to A_i^{\vphantom *} A_i^* \exp[ 2 \pi \I (\phi + \phi')]$.
These processes then lead to the magnetic flux dependencies given in the density
of states fluctuations, Eq.(\ref{eq4}).

From the density of states fluctuations the persistent
current fluctuations are found to be given by
\begin{eqnarray}
\langle I(\phi) I(\phi') \rangle &=& \sum_m {\cal C}_m 
\sin \left( { 2\pi m \phi \over \phi_0} \right) \sin 
\left({2\pi m \phi' \over \phi_0 } \right) \\
{\cal C}_m &= & { 8 e^2 m^2 L \over \pi^2 \hbar^2} \int_0^{\infty} {\D t \over t}
\left({ \pi k_B T \over \sinh( \pi k_B T t/\hbar) } \right)^2 P(mL,t)
\; .\end{eqnarray}
The temperature dependent factor originates from the Fourier transform of the Fermi function,
\begin{equation}
\int \D \epsilon f(\epsilon) \E^{\I \epsilon t/\hbar} 
= \I { \pi k_B T \over \sinh( \pi k_B T t/\hbar ) }
\; .\end{equation}
At zero temperature
only the diffusion time $\tau_d = L^2/D$ exists as a time scale in the above integral,
and we obtain
\begin{equation}
\langle I_{h/me}^2\rangle = {\cal C}_m= {96 \over \pi^2 m^3} \left({e \over \tau_d } \right)^2
.\end{equation}
At finite temperature there is an exponential cut-off for long paths, proportional to
$\exp( - 2 \pi k_B T t/\hbar)$,
due to the hyperbolic sine in the denominator.
This leads to a temperature dependence of the current on the diffusive scale
$k_B T\sim \hbar /\tau_d$ for the $h/e$ component, and 
$k_B T \sim \hbar/m^2\tau_d$ for the $m^{\rm th}$ Fourier component of the
current where the electrons have to diffuse $m$ times around the ring.
\section{INTERACTING ELECTRONS}
The Coulomb interaction enhances the average current considerably above the value for 
non-interacting electrons, as pointed out by Ambegaokar and Eckern\cite{ambegaokar90}.
Again we consider the grand canonical potential and work out the flux sensitive correction
in a simple approximation.
Consider the Hartree expression for the Coulomb interaction contribution to the thermodynamic
potential,
\begin{equation} \label{eqCoulomb}
\delta \Omega_C = {1\over 2} \int \D {\bf r} \D{\bf r}' v({\bf r} -{\bf r}')
\delta n({\bf r}) \delta n({\bf r}')
,\end{equation}
where $v({\bf r} - {\bf r}')$ is the screened interaction
and $\delta n({\bf r})$ is the spatially varying electron density.
In full analogy to what is found for the density of states,
the average electron density
is flux independent but the fluctuations do depend on magnetic flux.
Averaging the above equation with respect to disorder and including also the exchange (Fock) term,
we find
\begin{eqnarray}
\langle \delta  \Omega_C(\phi) \rangle 
&=& \mu_0 {2 L \over \pi \hbar} 
\int_0^\infty \D t \left({\pi k_B T \over \sinh( \pi k_B T t/\hbar) } \right)^2 \cr
\label{eqOinter}
&& 
\times \sum_m P(mL, t) \cos( 4\pi m \phi/\phi_0)
\; .\end{eqnarray}
The temperature dependent factor is related to the Fourier transform
of two Fermi functions, similar to the noninteracting theory. As above, 
the
diffusion probability $P(mL,t)$ appears. Since only time reversed 
paths contribute to the average current, the
primitive period is here given by $h/2e$, not $h/e$.
The dimensionless number $\mu_0$ characterizes 
the strength of the Coulomb interaction. It is given by the Fermi surface average
of the screened interaction, multiplied by the density of states.
For copper, for example, $\mu_0$ has been estimated to be about $0.3$.
The final result for the zero temperature persistent current is
$I \sim \mu_0 e/\tau_d$. For a precise estimate of the prefactor
the theory has to be refined: 
It is well known that higher order terms renormalize the coupling constant
$\mu_0$ logarithmically,
$\mu_0 \to \mu^* \approx \mu_0/\{ 1 + \mu_0  \ln[\epsilon_F/(\hbar/\tau_d)] \}$.
In addition, the attractive electron-phonon interaction reduces it even further.
This could, in principle, also lead to a different sign, implying that the system undergoes a 
transition to a superconducting state at extremely low temperature.

Taking $\mu^*$ as a parameter which has to be put into the theory ``by hand'',
we replace $\mu_0$ in Eq.\ (\ref{eqOinter}) by $\mu^*$, and find the
the zero temperature average persistent current to be given by
$I(\phi) = I_{h/2e} \sin( 4\pi \phi/\phi_0) + \dots $, with
$I_{h/2e} = 8 \mu^* /\pi (e/\tau_d)$.

\section{EXPERIMENTAL RESULTS}
Experimental results obtained by different groups are summarized in Table \ref{tab1}.
The mean free path $l$ and the phase coherence length $L_\phi$, given in the table,
have been determined by transport measurements in equally prepared wires.
In two of the experiments the number of rings was very large, $10^7$ and $10^5$ respectively. 
The current per ring should then correspond to the average current calculated in 
theory. 
In both experiments the $h/e$ component of the current was not observed, in agreement
with the theoretical prediction. Also the amplitude of the $h/2e$ component is reasonable,
both theory and experiment find a current of the order of $e/\tau_d$.
In the first experiment the sign of the current was not clear.
The amplitude of the experimental zero temperature current corresponds to an interaction
parameter $|\mu^*| = 0.3$, which is at least a factor five larger than what we expect
from the Coulomb interaction in copper.
In the second experiment,
Ref.\ 13,
the sign of the current was negative (diamagnetic),
which is clearly at odds with a theory taking into account the Coulomb
interaction only.
We emphasize that the computed temperature dependence is in
perfect agreement with the experimental findings of 
Ref.\ 2,
and in reasonable agreement
with 
Ref.\ 13:
Theory predicts in the relevant temperature range an exponential suppression,
$ I_{h/2e}(T) \approx I_{h/2e}(0) \exp( -T/T^*)$, with the typical temperature
given by $k_B T^* \approx 3 \hbar/\tau_d$. 
An exponential decay was seen in both experiments.
In the copper experiment\cite{levy90} the typical temperature was $T^*=80$ mK with
$\hbar/\tau_d \approx 25$ mK and in the semiconductor experiment $T^*=190$ mK
and $\hbar/\tau_d \approx 31$ mK.
\begin{table}
\begin{tabular}{|l|l|l|l|l|l|} \hline
                 & $L$& $l$& $L_\phi$   & $I_{h/e}$/ring        & $I_{h/2e}$/ring  \\
\hline \hline
$10^7$ rings; Cu & $2.2\mu$m  & $30$nm & $2\mu$m & $ 0$                  &  $\pm 0.77 e/\tau_d$ \\ 
(Ref. 2)                     &            &        &         &                       &                      \\\hline 
single rings; Au & $7.5\mu$m  & $70$nm & $12\mu$m  &   $+111 e/\tau_d $     & \\ 
(Ref. 3)                               & $12.5\mu$m & $70$nm & $12\mu$m  & $\pm 33e/\tau_d$ &\\ 
                 & $8.0\mu$m  & $70$nm & $12\mu$m  &    $+24e/\tau_d$ & \\ \hline
$30$ rings; Au   & $8\mu$m &  $87$nm & $16\mu$m &$ 0.42 e/\tau_d$& $- 0.44 e/\tau_d$         \\
(Refs. 11 and 12)                        &         &         &          &                & \\ \hline
$10^5$; GaAs/GaAlAs & $8\mu$m   &$3\mu$m   & $7\mu$m & $0$                & $-0.3 e/\tau_d$    \\
(Ref. 13)          &           &          &         &                    &                    \\ \hline
$16$; GaAs/GaAlAs   & $12\mu $m & $8\mu $m & $20\mu$m &$0.25 e/\tau_d$ &                 \\ 
(Ref. 14)           &           &          &         &                & \\\hline
\end{tabular}
\caption{Summary of experimental results for the persistent current in diffusive rings}
\label{tab1}
\end{table}

The experiments on single or a few rings are more difficult to interpret.
Theory predicts that in single rings the typical 
current is of the order $e/\tau_d$ and should vary from sample to sample.
The measured current in single gold rings\cite{chandrasekhar91}
was up to two orders of magnitude larger than this estimate.
More recent measurements of the current in a few rings, on the other hand, are much closer
to what is expected:
for $N$ rings the typical current per ring is expected to be of the order
$I \sim \langle I \rangle + \sqrt{ \langle \delta I^2 \rangle/N}$.
Measurements for 30 gold rings and 16 semiconductor rings are consistent with
this expectation, see Table \ref{tab1}. 
The measured $h/e$ current per ring, $I_{h/e} \approx 0.42 e/\tau_d$, is
close to the expected current, which is
of the order $\sqrt{96/ (30 \pi^2)} \approx 0.57 e/\tau_d$.
Concerning the $h/2e$ current it is not clear whether for $30$ rings the average
or the fluctuations dominate, since the expected fluctuations are
of the order $0.2 e/\tau_d$, which is only a factor two smaller than the experimentally
observed current.
The temperature suppression of the $h/e$ and the $h/2e$ current reported in 
Refs.\ 11 and 12
is consistent
with an exponential decay, 
with the characteristic temperature $T^*_{h/e} =166$ mK, and $T^*_{h/2e}=89$ mK.
The theoretical prediction in the relevant temperature range is\cite{cheung89}
$\langle  \delta I^2_{h/me} \rangle \approx \exp( -2 T/T^*_m)$, with
$k_B T^*_m= \pi^2 \hbar/(m^2 \tau_d)$, 
i.e. $T^*_{h/e} \approx 72$ mK and $T^*_{h/2e} \approx 18$ mK, when inserting 
$\hbar/\tau_d = 7.3$ mK.
The experimentally observed temperature suppression is thus somewhat slower than expected.

In conclusion we have seen that in some aspects theory and experiments agree with each other: 
The periodicity of the current is predicted correctly for both the experiments with
few and many rings.
The amplitude of the current is of the order $\hbar/\tau_d$, and the
temperature scale is reasonable, too.
Nevertheless the agreement is not satisfactory:
Although the $h/2e$ current is of the right order of magnitude, it is still
considerably larger than predicted. Moreover the negative
sign still requires an explanation.

Due to this discrepancies it is important to consider alternative mechanisms that
could influence the persistent current.
In the following we discuss recent suggestions which connect 
persistent currents and phase breaking.
\section{NOISE}
Recently it has been suggested\cite{mohanty99,kravtsov00} that
the large persistent currents might be related to another problem in mesoscopic physics,
namely the unexpectedly large electron dephasing rate.
Whereas it is expected that the dephasing rate decreases to zero in the low
temperature limit\cite{altshuler82}, many experiments show a saturation 
in this limit. 
Usually this saturation is attributed
to the presence of magnetic impurities or to heating.
At low temperature the latter is a serious problem,
since heating sets in already at very low applied voltages.
For example, when increasing the voltage, a saturation of the dephasing rate is seen
experimentally\cite{ovadyahu01},
even though the resistance continues
to increases with decreasing temperature.
However, it has been argued\cite{mohanty97}
that the above mentioned problems can be overcome, and that nevertheless a saturation of the 
dephasing rate is observed,

Several attempts have been made to explain this low temperature 
sa\-tu\-ra\-tion\cite{golubev98,altshuler98,imry99,zawadowski99}.
In particular it has been argued by Altshuler {\em et al.}\cite{altshuler98} 
that non-equilibrium electromagnetic noise can contribute to decoherence
without heating the electrons. 
Extending earlier work\cite{kravtsov93}  
on the effect of a high frequency electromagnetic field in mesoscopic rings,
Kravtsov and Altshuler\cite{kravtsov00}
have shown that non-equilibrium noise induces
a directed non-equilibrium current. This leads to the suggestion
that both the ``large'' currents 
ob\-served\cite{levy90,mohanty99,reulet95} and the strong dephasing
are related, and are non-equilibrium phenomena.

The starting point of the theory is the weak localization correction to the current
in the presence of magnetic flux and a time dependent electric field,
\begin{equation}
I_{wl}(t) = {{\cal C}_\beta e^2 D \over 2\hbar L}
\int_0^\infty \D \tau C_{t-\tau/2}\left({\tau\over 2}, -{\tau\over 2} \right) E(t-\tau)
\; ,\end{equation}
where $C_{t}(\tau, \tau')=\sum_q C_t(q, \tau, \tau')$ is the cooperon
at coinciding space points, and $ C_t(q, \tau, \tau')$  is determined by
\begin{equation}
{\partial C_t \over \partial \tau} + D 
\left(q - {e \over \hbar }(A_{t+\tau}+ A_{t-\tau}) \right)^2
C_t = \delta(\tau - \tau')
\; .\end{equation}
The momentum $q$ is for a ring structure given by $q= (2\pi/L)(n- 2\phi/\phi_0)$.
The constant ${\cal C}_\beta $ depends on the Dyson 
symmetry class: For pure potential disorder ($\beta=1$) ${\cal C}_\beta= -4/\pi$,
while ${\cal C}_\beta = 2/\pi$ in the presence of strong spin-orbit scattering ($\beta=4$).
Altshuler and Kravtsov calculated the $DC$-component of the current, i.e.
the time average $\overline{ I_{wl}(t)}$, in presence of a random electric field with
zero time average, $\bar A_t = 0$, and correlations $\overline{A_t A_{t'}}$ that decrease 
at $|t -t'| > t_c$, where $t_c  < \tau_\phi, \tau_d$.
The final result for the average persistent current is
\begin{equation}
I_{h/2me} = {\cal C}_\beta \left({ e \over \tau_\phi }\right) \exp\left( -m {L\over L_\phi} \right)
,\end{equation}
where
\begin{equation}
{1\over \tau_\phi } = 2 D(e^2/\hbar^2) \overline{ A_t^2}
\end{equation}
is the noise-induced dephasing rate.
If in one of the experiments dephasing is due to such a non-equilibrium
electric noise, then the noise-induced current is also relevant:
Since in all experiments the phase coherence length is of the order of the phase breaking length,
the noise induced current is of the order $e/\tau_d$.
The noise induced current is paramagnetic for rings with strong spin-orbit coupling (like 
in gold or copper), 
and is diamagnetic in the absence of spin-orbit coupling, and could thus be the
explanation of the semiconductor experiments in 
Ref.\ 13.
 
\section{IMPURITY MEDIATED INTERACTIONS}
In the presence of impurities, in particular, whenever the defects have an internal structure,
the persistent current may have a sizeable contribution through what can be called an
``effective interaction''; this mechanism was discussed, for example in 
Refs. 23 and 24.
We first 
consider the interaction of conduction electrons with nonmagnetic impurities, which we
assume to couple to the electron density. 
The Hamiltonian is of the form
\begin{equation}
\hat  H_{\rm int} =  \int \D x \hat n ({\bf x}) \hat V({\bf x}) 
\; .\end{equation}
The operator
$\hat V({\bf x})$ that is due to the impurities, will be specified more explicitly below.
To second order in this interaction one finds a correction to the free energy 
which is the sum of a Hartree and a Fock like term,
$\delta \Omega = \delta \Omega_{\rm H} + \delta \Omega_{\rm F}$.
We concentrate on the Hartree term
given by ($\beta = 1/k T $)
\begin{eqnarray}
\delta \Omega_{\rm H} &=& -{1\over 2 } \int_0^{\beta}\D \tau \int \D {\bf x} \int \D {\bf x'}
\langle \hat n ({\bf x} ) \rangle_{\rm th} \langle \hat n({\bf x'}) \rangle_{\rm th}
\nonumber \\
&&\label{hartree} \times \left[ \langle  \hat V({\bf x}, \tau) \hat V({\bf x'}, 0 ) \rangle_{\rm th} 
     - \langle \hat V({\bf x}) \rangle_{\rm th} \langle \hat V({\bf x'}) \rangle_{\rm th} \right] 
,\end{eqnarray}
where the brackets
$\langle \dots \rangle_{\rm th}$ are the thermal average.
Comparing Eqs.\ (\ref{eqCoulomb}) and (\ref{hartree}) 
one realizes that the Coulomb interaction is replaced by an 
effective interaction,  
\begin{eqnarray}
v({\bf x} - {\bf x'} )  \to - \int_0^\beta \D\tau 
\left[ \langle \hat V({\bf x}, \tau) \hat V({\bf x'} ,0) \rangle_{\rm th} -
\langle \hat V({\bf x}) \rangle_{\rm th} \langle \hat V({\bf x'}) \rangle_{\rm th} \right]
\end{eqnarray}
mediated by the defects.
If $\hat V({\bf x})$ describes pure potential scattering, 
then $\hat V({\bf x})$ is a c-number with the result that this effective interaction vanishes. 
The situation is more interesting whenever
the impurity has an internal degree of freedom.
For a two-level system, for example, 
which may be realized by an impurity which sits in a double well
potential with nearly degenerate minima at
${\bf r}$ and ${\bf r}+{\bf d}$,
we write the scattering potential as
$\hat V({\bf x}) = V [\hat n_A \delta({\bf x} -{\bf r} ) +  
                     \hat n_B \delta({\bf x} -{\bf r - d })]$. 
Here $\hat n_A$ and $\hat n_B$ are the number operators
for the impurity in the respective potential minimum. 
Since the impurity is in either of these minima,
$\hat n_A + \hat n_B = 1 $. 

Next we average the interaction over the Fermi surface.
The dimensionless interaction ``constant'', i.e. the analog to $\mu_0$ defined earlier for the
Coulomb interaction, reads
\begin{equation}\label{eq11}
\mu_{\rm TLS} = - {N_0 V^2 \over {\cal V} }{1\over 2} 
F
\int_0^\beta \D \tau  
\left[ \langle \hat n_A(\tau) \hat n_A(0) \rangle_{\rm th}- 
\langle\hat n_A \rangle_{\rm th} \langle \hat n_A \rangle_{\rm th} \right]
,\end{equation}
with
$F= \left[ 1- {\sin^2 (k_F d)/ (k_F d )^2 } \right]$, and ${\cal V}$ the volume.
In absence of spin-orbit scattering the
zero temperature persistent current thus is obtained as
\begin{equation}
I(\phi) = {16 \mu_{\rm TLS} \over \pi} {e\over \tau_d} \sin\left({4\pi \phi\over \phi_0} \right) + \cdots
.\end{equation}
In the presence of strong spin-orbit scattering this result has to be divided by four.
In order to calculate $\mu_{\rm TLS} $ explicitly, we 
need details of the impurity Hamiltonian.
We characterize the impurity by 
an asymmetry $\epsilon$, and a tunneling amplitude $\Delta$. 
The relevant correlation function 
is given by
\begin{equation} \label{eq13}
\int_0^\beta \D \tau
\left[
\langle \hat n_A(\tau) \hat n_A(0) \rangle_{\rm th} 
- \langle \hat n_A \rangle _{\rm th}\langle \hat n_A \rangle_{\rm th} \right]
= \left\{ \begin{array}{lc}
{1\over 4} {1\over k T } 
& \\ 
{1\over 4} {\Delta^2 \over \epsilon^2 + \Delta^2 }{1\over \sqrt{\epsilon^2 + \Delta^2}}
&   
\end{array}
\right.
\! \! \!  \! \! \! ,\end{equation}
in the two limits $\epsilon^2 + \Delta^2 < (kT)^2$, 
and $\epsilon^2 + \Delta^2 > (kT)^2$, respectively.
For a given concentration $c$ of two-level systems,
we find under the standard assumption\cite{black81}  
of a flat distribution of   
$\epsilon$ between zero and $\epsilon_{\rm max}$ ($> T$), 
and a distribution of $\Delta $ that
is proportional to $1/\Delta $ between $\Delta_{\rm min}$ and $\Delta_{\rm max}$,
the result
$\mu_{\rm TLS} \sim - F c N_0 V^2/\epsilon_{\rm max}$. 
This number should be of the order one, for this mechanism to be relevant for 
the persistent current experiments considered above. 
Estimates, however, 
are difficult due to the large number of parameters.
Although the required density of TLS is not unreasonable\cite{imry99,ahn01},
it has been objected\cite{ahn01,aleiner01} -- in another context -- 
that the required concentration is larger than
typical values in metallic glasses.

Certainly the estimate could be improved if some of the parameters could be determined
experimentally.
In 
Ref.\ 20
it has been demonstrated that in the presence of 
a sufficient number of
TLS, the electron dephasing rate becomes temperature independent in a
certain range of temperature.  
The dephasing rate is\cite{imry99} 
\begin{equation}
{1\over \tau_\phi} \sim \left\{ \begin{array}{ll} 
 \Delta_{\rm max}F c N_0 V^2/(\epsilon_{\rm max} \lambda  )  
& {\rm if}\,\, \hbar/\tau_\phi < \Delta_{\rm max} < k T\\
\Delta_{\rm max} ( F c N_0 V^2 /\hbar \lambda \epsilon_{\rm max} )^{1/2} 
& {\rm if }\,\,
\Delta_{\rm max}< \hbar/\tau_\phi  < k T\end{array}\right. 
\end{equation}
with $\lambda = \ln( \Delta_{\rm max}/ \Delta_{\rm min} ) $.
In an experiment where this mechanism is responsible for dephasing, 
we can express the persistent current amplitude,  
$I \sim \mu_{\rm TLS} (e/\tau_D)$, in terms
of the experimentally accessible dephasing rate as
\begin{equation}
|\mu_{\rm TLS}| \sim \left\{ \begin{array}{l}
 \lambda (\hbar/\tau_\phi)/ \Delta_{\max} \\
 \lambda (\hbar/\tau_\phi)^2/\Delta_{\rm max}^2
\end{array}\right.  
\end{equation}
in the two limits considered.
A possible candidate is the
gold sample of 
Ref.\ 11:
Below $500$ mK the dephasing rate is 
$T$-independent with $\hbar/\tau_\phi \sim 2$ mK.
For the mechanism considered here, the lowest measured
temperature ($\sim 40$ mK) is an upper limit for $\Delta_{\rm  max}$. This leads to the estimate
$|\mu_{\rm TLS}| > \lambda /20$.  
Note that the persistent current from TLS mediated interactions 
is diamagnetic, and could thus be responsible for
the diamagnetic $I_{h/2e}$ in the array of gold rings of Refs.\ 11 and 12.

\section{MAGNETIC DEFECTS}
Finally, it is of course important to consider magnetic impurities.
First, these are difficult to avoid in preparing the
experimental samples, and thus magnetic defects often have to be taken into account
for the interpretation of the results.
Second, however, magnetic impurities are well suited to change the properties
in an controlled way. 
Thus, in order to check the predictions given in this section, we would
strongly encourage such experiments.

\begin{figure}
\hspace{1.5cm}{\epsfxsize=8.5cm\epsfbox{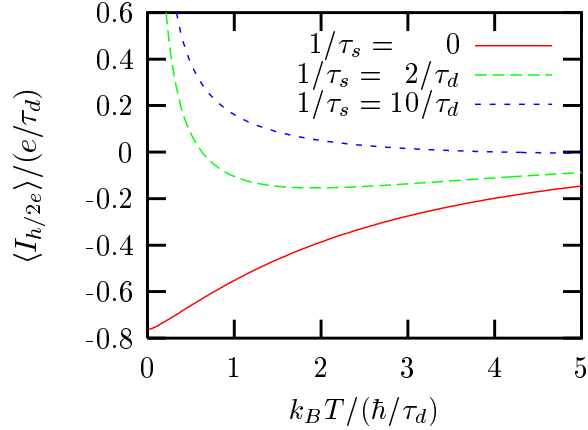}}
\caption{Prediction for the
average persistent current in the presence of magnetic impurities for
rings with strong spin-orbit scattering. The interaction parameter $\mu^*$ is chosen
$\mu^*= -0.3$.}
\label{fig2}
\end{figure}

The sensitivity of quantum coherence with respect to magnetic scattering is well known.
Magnetic impurities suppress quantum interference from time reversed paths. Therefore one
expects that the average persistent current in the presence of magnetic
impurities is suppressed; explicitly
\begin{eqnarray} 
\langle I(\phi) \rangle &= &  {4\mu^*e m L\over \pi \hbar^2}
\int_0^{\infty} \D t \left({ \pi k_B T \over \sinh( \pi k_B T t/\hbar)} \right)^2
\E^{-t/2 \tau_s} \cr
\label{eq24}
&& \times \sum_m P(mL, t) \sin( 4\pi m \phi/\phi_0)
\; ,\end{eqnarray}
where $\tau_s$ is the spin-flip scattering time. 
On the other hand the coupling of the conduction electrons with a local spin,
\begin{equation}
\hat H_{\rm int} = - J \hat{\bf  s}({\bf x}) \cdot \hat{\bf  S},
\end{equation}
will induce an effective, spin-dependent electron-electron interaction,
\begin{eqnarray}
v_{\alpha \beta \gamma \delta } ( \I \omega )&= &
- { J^2\over {\cal V}} 
\sum_{a,b=x,y,z} \sigma^a_{\alpha \gamma} \sigma^b_{\beta \delta}
\chi^{ab}(\I \omega) \\
\chi^{ab}(\I \omega) & =& 
\int_0^\beta \D \tau \E^{\I \omega \tau}
\left[ \langle \hat S^a(\tau) \hat S^b (0) \rangle_{\rm th} 
- \langle \hat S^a \rangle_{\rm th} \langle \hat S^b\rangle_{\rm th} \right]
\end{eqnarray}
which may
enhance the persistent current.
The effective interaction is proportional to the impurity susceptibility $\chi^{ab}$. 
This quantity, in weak magnetic fields, is proportional to the inverse temperature, at least when
Kondo correlations and spin-spin interactions are neglected.
\begin{figure}
\hspace{1.5cm}{\epsfxsize=8.5cm\epsfbox{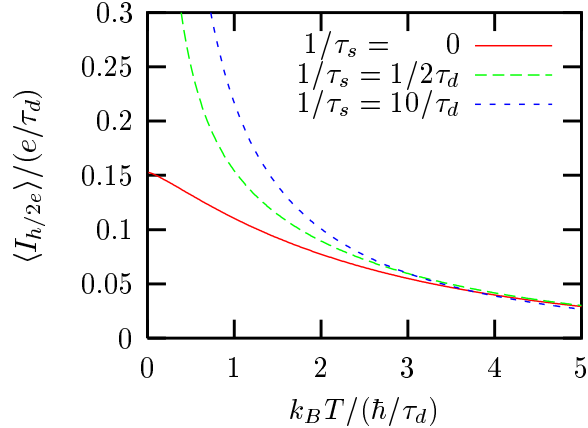}}
\caption{Average persistent current in the presence of magnetic impurities; here
the interaction parameter is 
$\mu^*=0.06$.}
\label{fig3}
\end{figure}
Figures \ref{fig2} and \ref{fig3} show the persistent current in the presence of the magnetic impurities and
for strong spin-orbit scattering, which is relevant for gold or copper.
The current was obtained by adding to Eq.\ (\ref{eq24}) the impurity induced current as
given in Ref.\ 23.
Notice the qualitative difference of the results for negative interaction parameter $\mu^*$ (Fig.\ \ref{fig2}),
compared to
positive $\mu^*$ (Fig.\ \ref{fig3}).
In the first case, a low concentration of magnetic impurities strongly suppresses the current,
and
at very low temperature there is a change in sign.
In the second case, where the persistent current in the 
absence of magnetic impurities is paramagnetic,
the current is not reduced by a low concentration of impurities; at low temperature it
is even enhanced considerably.
\section{SUMMARY}
One decade after the first experimental observation of
persistent currents in normal conducting rings the central question is still open:
Which mechanism is responsible for the amplitude of the current?
Existing theories capture correctly certain aspects of the experimental observations,
like the periodicity, the scale of the amplitude of the current and its temperature
dependence.
Nevertheless a theory of disordered electrons including the Coulomb interaction seems not
to be complete.

In this article we summarized recent ideas of the relationship between
persistent currents and dephasing.
All mechanisms considered have one
feature in common: If there is enough noise, 
(two-level systems, magnetic impurities) in order to explain the
experimentally observed dephasing, then the noise
is sufficient to explain the observed amplitude of the persistent current.

Experimentally the relation between dephasing and persistent currents may be checked
by measuring the persistent current for different materials.
For silver, where no saturation of the dephasing time has been observed\cite{gougam00},
we expect a smaller persistent
current than in gold or copper where the dephasing time saturates at low temperature.
We also suggest a study of the persistent current in samples doped with magnetic impurities.
By varying the impurity concentration one controls the strength of the impurity-mediated
electron-electron interaction. 
For the gold rings of Refs.\ 11 and 12
for example, we
predict a sign change of the $h/2e$ current with increasing concentration of magnetic impurities.

We acknowledge 
financial support by the Deutsche Forschungsgemeinschaft through SFB 484.

\end{document}